\documentclass[floatfix,twocolumn,showpacs,preprintnumbers]{revtex4}
\usepackage{graphicx}% Include figure files
\begin{document}

\title{Magnetocrystalline anisotropy and orbital polarization in ferromagnetic transition metals}

\author{Yuannan Xie}
\author{John A. Blackman}
\affiliation{Department of Physics, University of Reading,
Whiteknights, P. O. Box 220, Reading, RG6 6AF, United Kingdom.}

%\date{\today}

\begin{abstract}
The magnetocrystalline anisotropy energies (MAEs) of the ferromagnetic 
metals bcc Fe, fcc and hcp Co, and fcc Ni have been calculated by using 
the {\it ab initio} tight-binding method. Disentangling the strong 
correlation among the $d$ orbitals with the Hamiltonian in the local 
spin-density approximation, we have investigated the orbital polarizations
induced by the Hubbard $U$ and Racah $B$. The experimental MAE of fcc Ni 
is found with the value of $U$ close to that determined from experiments 
and used in other theories. With the optimized values of $U$ and 
$J$, both the MAEs and the orbital moments for Fe and Co are
in close agreement with experiment.    
\end{abstract}


\maketitle

Obtaining the magnetocrystalline anisotropy energies (MAEs) of Fe, Co, and
Ni from {\it ab initio} calculations within the local-spin-density
approximation (LSDA) to density functional theory is of considerable
current interest \cite{DKS90,TJEW95,HP98,B98,YSK01,S01}. From various
high-quality LSDA calculations, the best case is Fe, where the computed
values differ from experiment by a factor of about 2. The result for hcp Co
is far worse, and for Ni, the sign is not even correct. The effect of the
so called spin-other-orbit coupling is far too small to bring theory and
experiment into accord \cite{S01}. The discrepancy between theory and 
experiment, especially in the case of Ni, is usually attributed to the 
LSDA. 

The LSDA predicted MAEs and orbital moments can be improved for Fe and
Co \cite{TJEW95,DKS91} by introducing the Brooks' orbital
polarization (OPB) term \cite{Brooks85} which mimics Hund's second rule. 
However for Ni, the predicted easy axis is still wrong \cite{TJEW95}. 
In OPB, OP is driven by the Racah parameter $B$, with an energy 
functional related to the orbital moment $\langle\hat{L}\rangle$ given by
$\Delta E_{\text{OPB}} = -\frac{1}{2}B\langle\hat{L}\rangle^2$ \cite{EBJ90}. 
It was argued that the key parameter responsible for the
exchange-correlation enhancement of the orbital moments in solids
is the Hubbard $U$ rather than the intra-atomic Hund's second rule
coupling \cite{SLT98}. Recently, the experimental MAEs of Fe and Ni have 
been obtained\cite{YSK01} in the LDA$+U$ method \cite{AAL97} with  
the noncollinearity of intra-atomic magnetization included.
However, the authors found the MAE of fcc Ni to be a very rapidly
varying function of $U$ (from $-50$ to $60$ $\mu$eV/atom). 
A slight change of the value of $U$ ($\sim0.1$eV) may predict the wrong sign.
Given this sensitivity, it is highly desirable to disentangle the 
intra-atomic strong correlation with the Hamiltonian in the
LSDA and therefore to clarify the effect of OP in first principles 
calculations. This is the purpose of the present paper.

The basic features of the electronic structure of Fe, Co, and Ni can be
understood on the basis of their two types of valence-electron
orbitals \cite{SAS92}. The extended $s$, $p$ (,and $f$) orbitals should be
well described by the LSDA. The fairly localized $d$ orbitals, 
for which the electron-electron interaction is between the localized and
itinerant limits, are not adequately dealt with in the LSDA \cite{AAL97}. 
In order to disentangle the strong correlation with the LSDA, we express
the Hamiltonian as $\hat{H}=\hat{H}^{\text{LSDA}}+
\hat{H}^{\text{C}}+\hat{H}^{\text{SO}}$. Here $\hat{H}^{\text{LSDA}}$ is the
standard LSDA Hamiltonian, $\hat{H}^{\text{C}}$ is the correlation within the
$d$ orbital subspace and $\hat{H}^{\text{SO}}$ the spin-orbital interaction.

We start with the LSDA Hamiltonian in the orthogonal representation of
the tight-binding linear muffin-tin orbital method in the atomic sphere
approximation (TB-LMTO-ASA) \cite{A75},
\begin{equation}
\hat{H}^{\text{LSDA}} = C + \sqrt{\Delta}S^{\gamma}({\bf k})\sqrt{\Delta},
\label{e:hlmto}
\end{equation}
where $C$, $\Delta$, and $\gamma$ are the self-consistent standard potential
parameters. Because the electron-electron interaction is included in 
$\hat{H}^{\text{C}}$, the on-site diagonal matrix element of the $d$ 
orbital is replaced with $C_d=(C^{\uparrow}_d+C^{\downarrow}_d)/2$. 
$S^{\gamma}({\bf k})$ is the structure constant matrix in the orthogonal representation \cite{A75} with ${\bf k}$ running over the Brillouin zone (BZ). 
$\hat{H}^{\text{LSDA}}$ is block-diagonal in the spin index $\sigma$ along 
the magnetization direction. The spin-orbit coupling matrix elements for 
$d$ orbitals are calculated in the last iteration of the self-consistent 
field procedure \cite{DKS90} and $\hat{H}^{\text{SO}}$ is treated in the 
usual single-site approximation \cite{TBF76}.

Similarly to the LDA$+U$ method \cite{SLT98,AAL97}, we treat the screened 
interaction among the intra-atomic $d$-orbitals in the
Hatree-Fock approximation (HFA),
\begin{eqnarray}
E^{\text{ee}} = &\frac{1}{2}&\sum^{\sigma\sigma'}_{\{m\}}n^{\sigma}_{m_1m_2}
(U_{m_1m_3m_2m_4} \nonumber \\
&-&U_{m_1m_3m_4m_2}\delta_{\sigma,\sigma'})n^{\sigma'}_{m_3m_4}-E_a,
\label{e:einter}
\end{eqnarray}
where $U_{m_1m_3m_2m_4}=\langle m_1,m_3|V^{\text{ee}}|m_2,m_4\rangle$ and 
$n^{\sigma}_{m_1m_2}$ is the on-site $d$ occupation matrix in the
spin-orbital space. $E_a$ is the average interaction without spin and
orbital polarization. $U_{m_1m_3m_2m_4}$ are determined by three Slater
integrals $F_0$, $F_2$, and $F_4$\cite{T64}, which are linked to three
physical parameters: the on-site Coulomb repulsion $U=F_0$, exchange
$J =\frac{1}{14}(F_2+F_4)$, and Racah parameter
$B=\frac{1}{441}(9F_2-5F_4)$. In terms of $U$ and $J$, $E_a$ is expressed as
$\frac{1}{2}Un^2_d-\frac{U+4J}{5}(\frac{n_d}{2})^2$, where $n_d$ is the
on-site $d$-orbital occupation. The ratio $F_4/F_2$ is, to a good 
accuracy, a constant $\sim$0.625 for $d$ electrons \cite{AAL97}, which 
leads to the estimation $B\approx 0.11J$. The interaction energy 
$E^{\text{ee}}$, which is rotationally invariant with respect to the basis, 
leads to an effective potential $H^{{\text{C}}}$ acting on the 
$d$-orbital subspace.

The spin polarization (SP) in LSDA is generically close to the Stoner concept
with an energy related to spin magnetization $m$ of  
$\Delta E^{\text{LSDA}}_{\text{SP}}(m) = -\frac{1}{2}I(m)m^2$, where 
$I(m)$ is of the order of 1 eV\cite{G76}. In the HFA, the average spin 
splitting for $d$ electrons is driven by $I=\frac{1}{5}(U+4J)$, with 
OP determined by $U_{\text{eff}}=U-J$ \cite{B77}. 
In the limit $B=0$, $U_{m_1m_3m_2m_4}$ only involves two spherical
harmonics \cite{T64}, and $U=J=I$ is approximately equivalent to the
LSDA \cite{MM88}. Even with $U_{\text{eff}}=0$, there is no simple relation
between $E^{\text{ee}}$ and the orbital moment 
$\langle\hat{L}\rangle$\cite{SLT98}.
In practice, the problem involving the OPs induced by the 
Hubbard $U$ and Racah $B$ can be solved numerically by working directly
with the site-diagonal elements of the occupation matrix.  

The MAE is calculated by taking the difference of two total energies with
different directions of magnetization (MAE $= E_{111}-E_{001}$ for cubic
structures and MAE $= E_{10\bar{1}0}-E_{0001}$ for hcp structure). The total
energies are obtained via fully self-consistent solutions of $\hat{H}$, 
with the double counting corrections to the total energy included. 
For the ${\bf k}$-space integration, we use the special point
method \cite{F89} with a Gaussian broadening of 50 meV \cite{TJEW95}.
We use $100^3$ sampling points in the BZ for cubic structure and
$100\times100\times56$ points for hcp structure.
We have also included the occupation number broadening correction terms to 
the ground-state total energy \cite{GF98}. Numerical convergence has been 
tested against the number of ${\bf k}$-points and Gaussian broadening. 
We first calculate the electronic structure self-consistently using the 
scalar relativistic TB-LMTO-ASA method. Then we construct the 
Hamiltonian $\hat{H}$, using the spin-orbital coupling constants 
corresponding to the $d$ band center \cite{DKS90}. 

\begin{figure}[htbp]
\includegraphics[width=0.4\textwidth,angle=0]{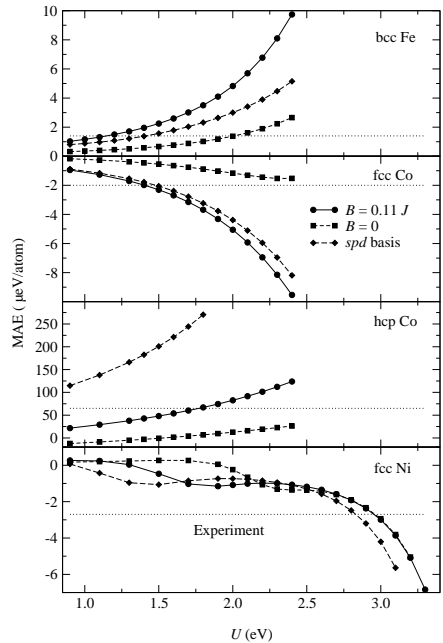}
\caption {MAEs of bcc Fe, fcc Co, hcp Co, and fcc Ni as a function 
of Hubbard $U$. Two curves in $spdf$ basis are plotted for 
each case, one with $B=0.11J$, the other with $B=0$. The experimental 
values are indicated by the horizontal dotted lines (1.4, 65, and 
$-2.7$ $\mu$eV/atom for bcc Fe, hcp Co and fcc Ni\cite{TJEW95}, and  
2.0 $\mu$eV/atom for fcc Co \cite{F98}). The MAEs calculated in $spd$ 
basis with $B=0.11J$ are also presented.}
\label{mfeconif1}
\end{figure}

\begin{figure}[htbp]
\includegraphics[width=0.4\textwidth,angle=0]{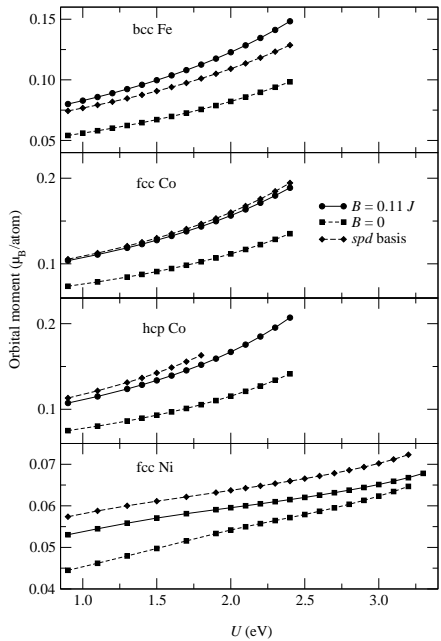}
\caption{Orbital magnetic moment for bcc Fe, fcc Co, hcp Co, and fcc Ni
along the experimental easy axis as a function of $U$. The plots are 
labelled the same as FIG. \ref{mfeconif1}.}
\label{mfeconif2}
\end{figure}

Considering the fact that the spin moments are well described by the LSDA, 
for a particular value of $U$, we have chosen the parameter $J$ such that
the magnetic moment maintains the theoretical value from the LSDA without
spin-orbit coupling. The calculated spin moments are almost independent of
$B$. Because the strong correlation $U$ and $J$ are entangled with the LSDA 
potential in the LDA$+U$ method \cite{AAL97}, the dependence of the magnetic
moment on $U$ and $J$ is not clear. Moreover, the energy $E^{\text{ee}}$ 
defined in the LDA$+U$ method \cite{SLT98,AAL97} is not zero even without 
any SP and OP. This may render the interpretation of the delicate
MAE dependence on $U$ quite difficult. In their LDA$+U$ calculations of Fe
and Ni, Yang et al. \cite{YSK01} scanned the ($U$,$J$) parameter space and
obtained the path of $U$ and $J$ values which hold constant the 
theoretical magnetic moment aligned along the (001) direction. $J$ 
increases with $U$ in their parameter path, in contrast to the basic 
concept $\frac{U+4J}{5}\approx I_d$. The SP and OP 
are treated on the same footing in our HFA scheme. The calculated MAEs
versus the values of $U$ are depicted in Fig. \ref{mfeconif1}.
The corresponding orbital moments are presented in Fig. \ref{mfeconif2}.     

%%%%%%%%%%%%%%%%Figure%%%%%%%%%%%%%%%%%%%%%%%%%%%%%%%%%%%%%%%%%%%%%%%%%%

%%%%%%%%%%%%%%%%%%%%%%%%%%%%%%%%%%%%%%%%%%%%%%%%%%%%%%%%%%%%%%%%%%%%%%%%%%

The ASA does not significantly affect the accuracy of MAE. In fact, the
differences in the MAEs and orbital moments calculated by the LMTO-ASA
method \cite{DKS90} and those calculated by the full potential (FP) LMTO
method \cite{TJEW95} are negligible if the same partial wave expansion
$l_{\text{max}}=2$ is used. In cubic structures, the difference between the
MAEs with $l_{\text{max}}=3$ and $l_{\text{max}}=2$ is very small 
($< 0.1$ $\mu$eV/atom)\cite{HP98}. However for hcp Co, the MAE changed 
sign when angular moment $l_{\text{max}}$ increased from 2 to 
3\cite{DKS90}. The MAE calculated in the $spdf$ LMTOs is closer 
to the recent accurate result calculated from FP linearized augmented 
plane-wave (LAPW) method \cite{S01}. Because the LMTOs in the $spdf$ 
basis are more complete than the ones in the $spd$ basis\cite{A75}, 
we regard the results in $spdf$ basis more reliable.  

In the limit $B=0$ and $U\approx J$, our calculated MAEs and orbital 
moments (the left ends of the $B=0$ plots) lie in the range of the recent 
high quality LSDA values \cite{DKS90,TJEW95,HP98,B98,YSK01,S01}. 
In particular, the predicted signs of the MAEs for hcp Co and fcc Ni are 
wrong in the $spdf$ basis. The calculated orbital moments are about 40\% 
smaller than the experimental values for Fe and Co, while for Ni, it is very 
close to the experimental value. Since the LSDA is even poorer in Ni than 
in Fe and Co\cite{SAS92}, the agreement must be accidental. 
 
By turning on the orbital polarization induced by the Racah $B$ at 
$U\approx J$, the calculated MAEs for Fe and Co are in much better 
agreement with the experimental data. This is similar to the LSDA+OPB 
calculations\cite{TJEW95,DKS91}. Particularly, the predicted easy axis 
is correct for hcp Co \cite{DKS91}. The calculated MAE in $spdf$ basis for 
hcp Co is quite close to that of the LMTO-ASA calculation with 
OPB\cite{DKS91}, while our $spd$ MAE is quite close to the FP-LMTO$+$OPB 
result\cite{TJEW95}. Interestingly, as shown in Fig. {\ref{mfeconif2}}, 
the enhancement of orbital moment due to OP induced by Racah 
$B$ is in excellent agreement with the OPB calculations \cite{TJEW95} 
despite the forms of $E^{\text{ee}}$ and $\Delta E_{\text{OPB}}$ being  
quite different. We suggest that the widely used OPB can be brought 
{\it precisely into accord with} the unrestricted HFA with 
$\Delta E_{\text{OPB}}$ replaced by $E^{\text{ee}}$ at $U=J$ 
($\frac{n_d}{2}$ in $E_a$ replaced with $n^{\uparrow}_d$ and 
$n^{\downarrow}_d$). Only with the OP induced by $B$, 
the predicted sign of MAE for fcc Ni is wrong.

We now study the effect of OP induced by $U$ with $U_{\text{eff}}>0$. 
Similarly to the OP induced by $B$, as shown in Fig. \ref{mfeconif2}, the 
OP induced by $U$ enhances the orbital magnetic moments. For both Fe and Co, 
the MAEs change very smoothly and monotonically with increasing $U$. For Ni, 
the MAE is about zero when $U\approx 1.3$ eV. It decreases with
increasing $U$, then making a flat region with MAE$\approx-1$ eV from
$U=1.7$ to $U=2.5$. After the flat region, the MAE decreases with increasing 
$U$, reaching the experimental value at $U=2.95$ eV. The MAE predicted here
changes very smoothly with increasing $U$ and with correct sign when
$U>1.3$ eV, without the strong sensitivity observed in the LDA$+U$
calculations \cite{YSK01}.
 
\begin{table}[htbp]
\caption{$U$ and $J$ (in eV) corresponding to experimental magnetic
anisotropy energy (in $spdf$ basis with $B=0.11J$). The calculated orbital
moments $l_z$ ($\mu_{\text{B}}$/atom) along easy axis are compared with the
experimental data \cite{TJEW95}.}
\label{mfeconit1}
\begin{ruledtabular}
\begin{tabular}{ccccc}
 & bcc Fe & fcc Co & hcp Co & fcc Ni \\
\hline
$U$ & 1.15 & 1.41 & 1.77 & 2.95 \\
$J$ & 0.97 & 0.83 & 0.75 & 0.28 \\
$l_z$ & 0.087 & 0.123 & 0.150 & 0.064 \\
expt. & 0.08  &       & 0.14  & 0.05  \\
\end{tabular}
\end{ruledtabular}
\end{table}

When $U>2.5$ eV, the two MAE curves of $B=0$ and $B=0.11J$ are almost 
indistinguishable for fcc Ni. Thus we conclude that it is the Hubbard 
$U$ that is fully responsible for bringing theory into accord with 
experiment.
For Fe and Co, both OPs induced by $U$ and $B$ are needed to produce the
experimental MAEs. As shown in Table \ref{mfeconit1}, for bcc Fe and hcp
Co, the optimized $U$ and $J$ almost simultaneously give the experimental
MAEs and orbital moments. The predicted orbital moment for fcc Ni is
slightly higher, but quite acceptable. The optimized values of $U$ and $J$
and their trend from Fe to Ni are very similar to those determined from
experiments and used in other theories \cite{SAS92,LKK01}. The optimized
values of $U$ for fcc Co and hcp Co are close but not the same. This may be
due to the fact that the experimental MAE of fcc Co was extracted 
from measurement on supported films \cite{F98} or that the calculated MAE 
of hcp Co is not fully converged even with $l_{\text{max}}=3$.  
 
%%%%%%%%%%%%%%%%%%Figure%%%%%%%%%%%%%%%%%%%%%%%%%%%%%%%%%%%%%%%%%%%%%%%%%%
\begin{figure}[htbp]
\includegraphics[width=0.35\textwidth,angle=270]{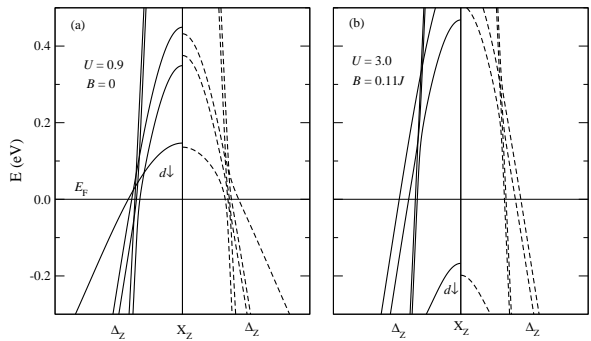}
\caption{Band structure of fcc Ni along $\Delta_Z=2\pi/a(0,0,l)$ with the
spin quantization axis in the (001) (solid lines) and the (111)
(dashed lines) directions. Here $0.4\leq l\leq 1.0$. $l=1.0$ corresponds to
the $X_Z$ point.}
\label{mfeconif3}
\end{figure}
%%%%%%%%%%%%%%%%%%%%%%%%%%%%%%%%%%%%%%%%%%%%%%%%%%%%%%%%%%%%%%%%%%%%%%%%%%

It was conjectured that the failure of the LSDA to predict the MAE of 
fcc Ni is related to the band structure along $\Gamma X$ 
direction \cite{DKS90}. As shown in Fig.\ref{mfeconif3}, similarly to 
the LSDA bands\cite{DKS90}, five bands cross the Fermi energy almost at 
the same ${\bf k}$ point when $J\approx U=0.9$ eV and $B=0$ (The bands for 
$B=0.11J$ are quite similar). One of the $d$ bands is just above the 
Fermi level at the $X$ point and results in the appearance of small 
$X_2$ pocket on the Fermi surface\cite{DKS90,YSK01}, which has not been 
found experimentally. Increasing the valence electrons and thus pushing 
down the band corresponding to the $X_2$ pocket, Daalderop 
et al. \cite{DKS90} found the correct easy axis for fcc Ni. We find that 
the $X_2$ pocket disappears at $U\approx 1.5$ eV, corresponding to the 
start point of the correct sign of the MAE. With increasing $U$, this band
is further pushed down and when $U\approx 3$ eV, it is below the Fermi
energy by about 0.2 eV. In our approach, the disappearance of the $X_2$ 
pocket is a natural result of the OP induced by Hubbard $U$. 

As a delicate property, the MAE is naturally expected to depend on the 
delicate changes of the band structures. We have applied the current scheme
to the parametrized TB models of Fe and Ni fitting to the APW bands in the
LSDA \cite{DAP86}. The trend and optimized values of $U$ and $J$ are found
to be very similar to the {\it ab initio} TB calculations. This underlines
the importance of how to treat the intra-atomic strong correlation. 

In summary, we have calculated the MAEs of Fe, Co, and Ni from the   
{\it ab initio} tight-binding total energies. Disentangling the strong 
correlation among the intra-atomic $d$ orbitals with the 
Hamiltonian in the LSDA and therefore treating the SP 
and OP on the same footing, we have solved the 
long-standing notorious problem of the MAE of fcc Ni. The discussions  
on the OPs induced by Hubbard $U$ and Racah $B$ can shed light on 
future first principles and model TB calculations. How to calculate 
the interaction parameter $U$ in metallic environment directly from 
first principles is still an open problem.          

This work was supported by the EU through the AMMARE project
(Contract No. G5RD-CT-2001-00478) under the Competitive and
Sustainable Growth Programme.

%\begin{references}

%\end{references}
%%%%%%%%%%%%%%%%%%%%%%%%%%%%%%%%%%%%%%%%%%%%%%%%%%%%%%%%%%%%%%%%%%%%%%%%
%\newpage
%\begin{figure}[htbp]
%\includegraphics[width=0.8\textwidth,angle=0]{mfeconi1.ps}
%\caption {MAEs of bcc Fe, fcc Co, hcp Co, and fcc Ni as a function 
%of Hubbard $U$. Two curves in $spdf$ basis are plotted for 
%each case, one with $B=0.11J$, the other with $B=0$. The experimental 
%values are indicated by the horizontal dotted lines (1.4, 65, and 
%$-2.7$ $\mu$eV/atom for bcc Fe, hcp Co and fcc Ni\cite{TJEW95}, and  
%2.0 $\mu$eV/atom for fcc Co \cite{F98}). The MAEs calculated in $spd$ 
%basis with $B=0.11J$ are also presented.}
%\label{mfeconif1}
%\end{figure}

%\newpage
%\begin{figure}[htbp]
%\includegraphics[width=0.8\textwidth,angle=0]{mfeconi2.ps}
%\caption{Orbital magnetic moment for bcc Fe, fcc Co, hcp Co, and fcc Ni
%along the experimental easy axis as a function of $U$. The plots are 
%labelled the same as FIG. \ref{mfeconif1}.}
%\label{mfeconif2}
%\end{figure}

%\newpage
%\begin{figure}[htbp]
%\includegraphics[width=0.7\textwidth,angle=270]{mfeconi3.ps}
%\caption{Band structure of fcc Ni along $\Delta_Z=2\pi/a(0,0,l)$ with the
%spin quantization axis in the (001) (solid lines) and the (111)
%(dashed lines) directions. Here $0.4\leq l\leq 1.0$. $l=1.0$ corresponds to
%the $X_Z$ point.}
%\label{mfeconif3}
%\end{figure}
%%%%%%%%%%%%%%%%%%%%%%%%%%%%%%%%%%%%%%%%%%%%%%%%%%%%%%%%%%%%%%%%%%%%%%%%%%%%%


\begin{thebibliography}{26}

\bibitem{DKS90}
G.H.O. Daalderop {\it et al.}, Phys. Rev. B {\bf41}, 11919 (1990).

\bibitem{TJEW95}
J. Trygg {\it et al.}, Phys. Rev. Lett. {\bf75}, 2871 (1995), and
references therein.

\bibitem{HP98}
S. V. Halilov {\it et al.}, Phys. Rev. B {\bf57}, 9557 (1998).

\bibitem{B98}
S. V. Beiden {\it et al.}, Phys. Rev. B {\bf57}, 14247 (1998).

\bibitem{YSK01}
I. Yang {\it et al.}, Phys. Rev. Lett. {\bf87},
216 405 (2001).

\bibitem{S01}
M. D. Stiles {\it etal.}, Phys. Rev. B {\bf64}, 104430 (2001).

\bibitem{DKS91}
G.H.O. Daalderop {\it et al.}, Phys. Rev. B {\bf44}, 12054 (1991).

\bibitem{Brooks85}
M. S. S. Brooks, Physica B {\bf130}, 6 (1985).

\bibitem{EBJ90}
O. Eriksson {\it et al.}, Phys. Rev. B {\bf41}, 7311 (1990).

\bibitem{SLT98}
I.V. Solovyev {\it et al.}, Phys. Rev. Lett. {\bf80}, 5758 (1998).

\bibitem{AAL97}
V.I. Anisimov {\it et al.}, J. Phys.: Condens. Matter {\bf9}, 767 (1997).

\bibitem{SAS92}
M. M. Steiner {\it et al.}, Phys. Rev. B {\bf45}, 13 272 (1992), and
references therein.

\bibitem{A75}
O.K. Andersen, Phys. Rev. B {\bf12}, 3060 (1975); O.K. Andersen
{\it et al.}, Phys. Rev. B {\bf34}, 5253 (1986).

\bibitem{TBF76}
H. Takayama {\it et al.}, Phys. Rev. B {\bf 14}, 2287 (1976).

\bibitem{T64}
M. Tinkham, {\it Group Theory and Quantum Mechanics}
(McGraw-Hill, New York, 1964), P. 174.

\bibitem{G76}
O. Gunnarsson, J. Phys. F {\bf 6}, 587 (1976).

\bibitem{B77}
B. Brandow, Adv. Phys. {\bf 26}, 651 (1977).

\bibitem{MM88}
It is equavalent to the LSDA if $I_d(m)$ in the LSDA is independent of 
the magnetization $m$. In fact, $I_d(m)$ depends on $m$ very weakly, e.g.,   
P.M. Marcus and V. L. Moruzzi, Phys. Rev. B {\bf 38}, 6949 (1988).

\bibitem{F89}
S. Froyen, Phys. Rev. B {\bf 39}, 3168 (1989).

\bibitem{GF98}
O. Grotheer and M. F\"{ahnle}, Phys. Rev. B {\bf 58}, 13459 (1998).

\bibitem{F98}
J. Fassbender {\it et al.}, Phys. Rev. B {\bf 57}, 5870 (1998).

\bibitem{LKK01}
A.I. Lichtenstein {\it et al.}, Phys. Rev. Lett. {\bf 87}, 67205 (2001).

\bibitem{DAP86}
D.A. Papaconstantopoulos, {\it Handbook of the Band Structure of 
Elemental Solids} (Plenum, New York, 1986) 

\end{thebibliography}
\end{document}